\documentclass[conference]{IEEEtran}
\IEEEoverridecommandlockouts
\usepackage{cite}
\usepackage{amsmath,amssymb,amsfonts}
\usepackage{algorithmic}
\usepackage{graphicx}
\usepackage{soul}
\usepackage{textcomp}
\usepackage{xcolor}
\usepackage{multirow}
\usepackage{makecell}
\def\BibTeX{{\rm B\kern-.05em{\sc i\kern-.025em b}\kern-.08em
    T\kern-.1667em\lower.7ex\hbox{E}\kern-.125emX}}
\usepackage{listings}
\usepackage{tcolorbox}
\usepackage{hyperref} 

\definecolor{promptbg}{RGB}{200,200,200}
\definecolor{responsebg}{RGB}{215,230,250}

\begin{document}

\IEEEaftertitletext{\vspace{-2.0\baselineskip}}

\title{PyraNet: A Multi-Layered Hierarchical Dataset for Verilog \\ 
}

\author{\IEEEauthorblockN{1\textsuperscript{st} Bardia Nadimi}
\IEEEauthorblockA{\textit{Dept. of Computer Science \& Eng.} \\
\textit{University of South Florida}\\
Tampa, Florida, United States \\
bnadimi@usf.edu}
\and
\IEEEauthorblockN{2\textsuperscript{nd} Ghali Omar Boutaib}
\IEEEauthorblockA{\textit{Dept. of Computer Science \& Eng.} \\
\textit{University of South Florida}\\
Tampa, Florida, United States \\
ghaliomar@usf.edu}
\and
\IEEEauthorblockN{3\textsuperscript{rd} Hao Zheng}
\IEEEauthorblockA{\textit{Dept. of Computer Science \& Eng.} \\
\textit{University of South Florida}\\
Tampa, Florida, United States \\
haozheng@usf.edu}
}

\maketitle

\begin{abstract}
Recently, there has been a growing interest in leveraging Large Language Models for Verilog code generation. 
However, the current quality of the generated Verilog code remains suboptimal. 
This is largely due to the absence of well-defined, well-organized datasets with high-quality samples, as well as a lack of innovative fine-tuning methods and models specifically trained on Verilog. 
In this paper, we introduce a novel open-source dataset and a corresponding fine-tuning technique, which utilizes a multi-layered structure that we refer to as PyraNet. 
Our experiments demonstrate that employing the proposed dataset and fine-tuning approach leads to a more accurate fine-tuned model, producing syntactically and functionally correct Verilog code.
The evaluation results show improvements by up-to $32.6\%$ in comparison to the CodeLlama-7B baseline model and up-to $16.7\%$ in comparison to the state-of-the-art models using VerilogEval evaluation platform.
\end{abstract}


\vspace*{-8pt}
\section{Introduction and Motivation}

The introduction of attention-based models~\cite{attentionIsAllYouNeed} represents a landmark development in the field of language processing. 
This breakthrough spurred the widespread adoption of transformer architectures, driving significant advancements in the area. 
Many models, such as Generative Pre-trained Transformers (GPT)\cite{radford2018GPT}, Bidirectional Encoder Representations from Transformers (BERT)\cite{BERT}, and Language Model for Dialogue Applications (LaMDA)~\cite{LaMDA}, have leveraged transformer technology to achieve notable success.

The motivation behind utilizing large language models (LLMs) for hardware code generation is centered on developing a tool that streamlines the hardware modeling process for designers \cite{MaskedLM ,spicepilot}. 
It also enhances the security aspects of hardware design \cite{Navabi}.
Integrating LLMs into this field aims to reduce the complexities traditionally involved in hardware design. 
By harnessing the advanced features of these models, hardware model development is expected to become more intuitive and efficient.

Additionally, the automation of hardware code generation through LLMs is vital for reducing human error. 
Given the intricate and technical nature of hardware design, manual coding is often prone to mistakes \cite{Hardfails}. 
By automating the process, LLMs not only speed up development but also significantly lower the chances of introducing errors that could arise from human oversight. 
This results in more reliable and robust hardware designs, as the automated system ensures consistent, high-quality code generation with reduced risk of faults. 
Ultimately, employing LLMs in this domain marks a move toward more efficient, error-resilient, and user-friendly approaches to hardware design.

While there has been extensive research focused on software program synthesis, hardware code synthesis remains relatively underexplored. 
Recently, researchers have begun to investigate the use of LLM architectures for generating Hardware Description Language (HDL) code~\cite{BenchmarkingVerilog, VerilogEval, ChipGPT, RTLLM, ChipChat, GPT4AIGchip, ChipNeMo, RTLFixer, AutoChip, rtlcoder, improvingLLMforHDL}.

Despite early successes in using LLMs for HDL code generation in hardware design, several challenges remain that must be addressed to push the field forward. 
One key issue is the limited availability of labeled data necessary for effective fine-tuning of LLMs, as seen in the more mature field of software code generation. 
Existing approaches to Verilog code generation often suffer from syntax and functionality errors~\cite{BenchmarkingVerilog, VerilogEval}, highlighting the need for further improvements.

In this paper, we introduce PyraNet, a multi-layered open-source dataset structure \footnote{\href{https://huggingface.co/datasets/bnadimi/PyraNet-Verilog}{https://huggingface.co/datasets/bnadimi/PyraNet-Verilog}} that utilizes a wide range of open-source Verilog samples, irrespective of their code quality. By organizing the dataset into different quality tiers, we are able to maximize the value from both the high-quality samples and those of lower quality. The overall structure of the dataset is depicted in Fig.\ref{fig:overallFineTuningArchitecture}-a.

The proposed fine-tuning technique is designed to fully leverage all data entries across the different tiers of the dataset. 
This is done by assigning higher loss weights to the high-quality data, and progressively reducing these weights for lower-quality data as we move down the pyramid \cite{lossWeighting}. 
Additionally, we implement a curriculum learning approach \cite{curriculumLearning}, starting the fine-tuning process with simpler data and gradually introducing more complex data as training progresses. 
By combining both the loss weighting and curriculum learning strategies, we can achieve a more effective fine-tuned model.
The overall proposed fine-tuning architecture is depicted in Fig.~\ref{fig:overallFineTuningArchitecture}.
The \textbf{key contributions} of this paper is summerized as follows:

\begin{figure*}[t]
    \centering
    \includegraphics[width=0.98\textwidth]{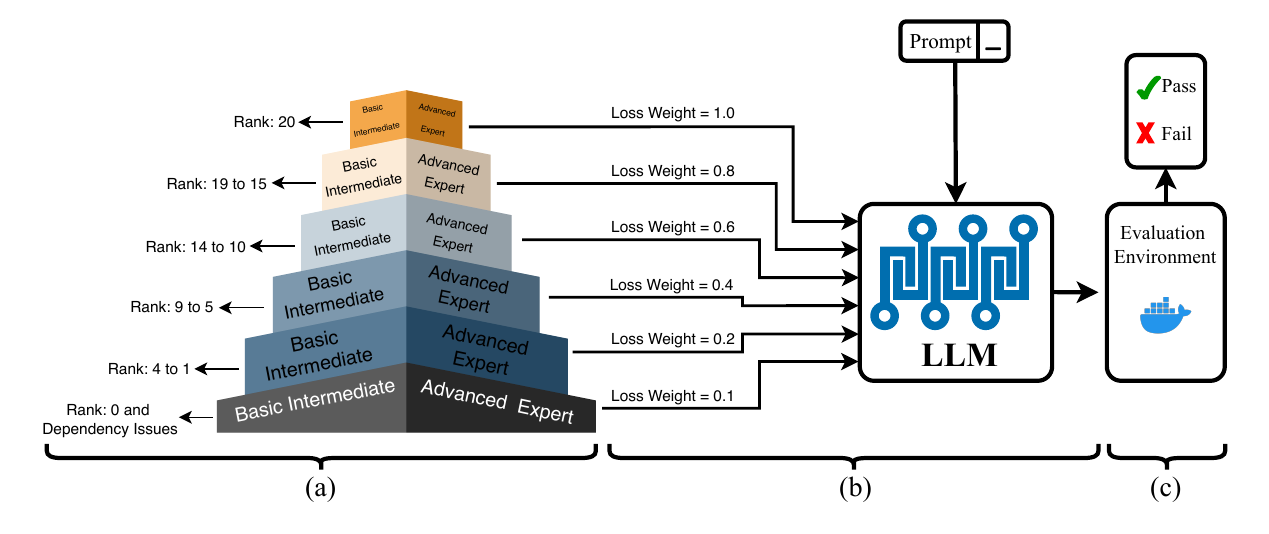}
    \vspace*{-14pt}
    \caption{Proposed Overall Architecture. (a) PyraNet dataset architecture. (b) PyraNet fine-tuning architecture. (b) Evaluation process.}
    \label{fig:overallFineTuningArchitecture}
    \vspace*{-14pt}
\end{figure*}

\begin{itemize}
    \item The introduction of the PyraNet open-source dataset that includes various quality tiers, categorized based on the quality of the Verilog code, as well as the available labels for each entry. The labels include information such as the complexity level of the code, code rankings, design descriptions, and compile details.
    \item The introduction of the corresponding fine-tuning approach which is designed to maximize the benefits of the proposed PyraNet dataset by incorporating both the loss weighting technique and the curriculum learning strategy.
    \item Experiments show that the new PyraNet approach improves the Verilog code generation by up-to $32.6\%$ in comparison to the CodeLlama-7B baseline model and up-to $16.7\%$ in comparison to the state-of-the-art approaches including RTLCoder \cite{rtlcoder} and OriGen \cite{origen}.
\end{itemize}

The remainder of this paper is organized as follows: Section II reviews related research in the field. Section III introduces the methodology, along with the proposed dataset and fine-tuning approach. Section IV focuses on comparisons and evaluations. Lastly, conclusions and future directions are discussed in Section V.
\vspace*{-6pt}

\vspace*{-6pt}
\section{Related Works}
\vspace*{-6pt}

The study in \cite{BenchmarkingVerilog} explores the capability of LLMs to generate Verilog code for hardware design, demonstrating significant success in achieving syntactically accurate outputs through fine-tuning on Verilog-specific datasets. By emphasizing the role of LLMs in reducing human errors and enhancing automation in hardware design, the research underscores the need for further advancements in functional correctness and sets the stage for broader AI integration in hardware development.

The study in~\cite{VerilogEval} presents VerilogEval, a benchmarking framework for evaluating the performance of LLMs in generating Verilog code for hardware design and verification. Utilizing a dataset of 156 problems from HDLBits, it investigates the automation of Verilog code generation across varying complexities, from simple circuits to advanced finite-state machines. The findings demonstrate that fine-tuning LLMs improves the quality of generated code, highlighting the transformative role of AI in optimizing hardware design workflows. Additionally, the research emphasizes the potential of supervised fine-tuning to enhance LLM capabilities, representing a significant step forward for both academic research and practical applications in the domain.

The study in~\cite{RTLLM} addresses the absence of standardized benchmarks for evaluating LLMs, such as ChatGPT, in hardware design, with a focus on Register Transfer Level (RTL) generation from natural language. It introduces RTLLM, an open-source benchmark designed to evaluate the effectiveness of LLMs in RTL design, establishing new standards for syntax, functionality, and design quality. The study also proposes a novel prompt engineering technique, "self-planning," which enhances GPT-3.5's performance, representing a major advancement in leveraging LLMs for complex and scalable hardware design tasks.

MG-Verilog \cite{mgverilog} introduces a multi-grained dataset tailored for LLM-assisted Verilog code generation, addressing limitations in size, complexity, and description granularity found in existing hardware datasets. MG-Verilog includes over 11,000 Verilog code samples with corresponding descriptions at varying levels of detail, such as high-level summaries, block summaries, and line-by-line comments. The dataset facilitates balanced fine-tuning, enabling LLMs to improve both code implementation accuracy and generalization across diverse hardware design tasks. Extensive experiments demonstrate the dataset's effectiveness, showcasing superior performance compared to models trained on other datasets, with enhanced metrics in code generation accuracy and functionality. 

RTLCoder \cite{rtlcoder} introduces a fully open-source framework for RTL code generation, addressing the challenges of privacy concerns and the lack of high-performance open-source LLMs for Verilog generation. The framework includes a dataset of over 27,000 instruction-code pairs, generated through an automated pipeline that ensures diversity and complexity across a wide range of hardware design tasks. RTLCoder employs a lightweight LLM architecture with only 7 billion parameters, fine-tuned using a novel training scheme that incorporates code quality feedback to enhance generation accuracy.

MEV-LLM \cite{Nadimi2024-MEV-LLM} presents a multi-expert large language model architecture tailored for Verilog code generation, addressing challenges of syntax and functionality errors in existing approaches. MEV-LLM employs multiple expert models, each fine-tuned to handle specific design complexity levels, ranging from basic to expert-level Verilog designs. The study develops a categorized dataset with fine-grained and coarse-grained labels to enhance the fine-tuning process, enabling better alignment between input descriptions and generated code. Evaluation against state-of-the-art models demonstrates significant improvements, with MEV-LLM achieving up to a 23.9\% increase in the pass@k metric. 

The paper \cite{origen} introduces OriGen, an open-source framework for Register Transfer Level (RTL) code generation, designed to address the limitations of existing approaches in data quality and self-reflection. OriGen employs a novel code-to-code augmentation methodology to generate high-quality RTL datasets and incorporates a self-reflection mechanism for error correction using compiler feedback. The framework trains two specialized LoRA (Low-Rank Adaptation) models, one for generating initial RTL code and another for rectifying syntactic errors, enhancing its ability to handle complex Verilog tasks. Experimental results show that OriGen outperforms all existing open-source LLMs, achieving a 12.8\% improvement over the best-performing model on the VerilogEval-Human benchmark and demonstrating superior self-reflection capabilities, surpassing GPT-4 Turbo by 19.9\% in syntactic correctness. 

CodeV \cite{CodeV} presents a robust framework for Verilog code generation by leveraging multi-level summarization and fine-tuning large language models (LLMs). The approach constructs a high-quality dataset of 165,000 Verilog modules collected from real-world sources, enriched by GPT-3.5-generated functional and high-level summaries. The experimental results demonstrate state-of-the-art performance on benchmarks like VerilogEval and RTLLM, with CodeV surpassing GPT-4 by 22.1\% on VerilogEval and achieving notable gains in syntax and functional correctness. This work highlights the potential of combining domain-specific datasets and advanced fine-tuning techniques to push the boundaries of hardware description language generation.

BetterV \cite{BetterV} introduces a framework for controlled Verilog generation that enhances Verilog implementation through a novel combination of instruct-tuning and generative discriminators. BetterV utilizes a domain-specific dataset, created through rigorous filtering and augmentation techniques, and aligns Verilog with C programs to enhance large language models' (LLMs) understanding of hardware description languages. By employing task-specific generative discriminators, BetterV optimizes downstream Electronic Design Automation (EDA) tasks such as synthesis node reduction and verification runtime. The framework achieves state-of-the-art performance on the VerilogEval benchmark, outperforming GPT-4 in functional correctness. 

AutoVCoder \cite{AutoVCoder} addresses challenges in register-transfer level (RTL) code generation through three key innovations: a high-quality hardware dataset generation strategy, a two-round fine-tuning process, and a domain-specific retrieval-augmented generation (RAG) mechanism. Experimental evaluations demonstrate AutoVCoder's superiority over existing state-of-the-art models, achieving notable improvements across multiple benchmarks, including up to 3.4\% gains in functional correctness on the RTLLM benchmark. By integrating domain-specific retrieval and fine-tuning techniques, AutoVCoder bridges the gap between general-purpose LLM capabilities and the specialized requirements of RTL code generation, paving the way for efficient and high-quality hardware design automation.

Each of the previously discussed studies faces challenges in producing Verilog codes that are both syntactically and functionally accurate.
\vspace*{-6pt}
\section{Methodology and Dataset}
\vspace*{-4pt}
This section starts by explaining the structure and organization of the PyraNet dataset, followed by a detailed description of the proposed fine-tuning process.
\vspace*{-6pt}
\subsection{PyraNet Dataset}

\subsubsection{Dataset gathering}
The majority of our dataset was gathered from publicly accessible GitHub repositories. 
Additionally, some Verilog code examples were generated using commercial large language models like GPT-4o-mini.

\subsubsection{Post-download dataset filters}
Following the collection of all Verilog code samples, we implemented multiple filtering procedures to guarantee the dataset's quality and relevance. The following are the filtration steps:
\begin{itemize}
    \item {\tt Empty/Broken files:} The initial step involved the removal of empty and corrupted or broken files. A Python script was employed to detect files with encoding issues by attempting to read each file. Any files that could not be processed due to encoding errors were discarded. Likewise, files that were successfully read but were empty were also excluded from the dataset.
    
    \item {\tt Module declaration:} Files without a valid module declaration were similarly filtered out. Using a Python script, we identified and excluded any files that lacked module declarations, following the same procedure as the check for empty or corrupted files.
    \item {\tt Code Deduplication:} We employed the Jaccard similarity algorithm to perform deduplication. This method computes the similarity between sets of tokens derived from the code samples by measuring the intersection over the union of the sets. Code pairs with a Jaccard similarity score above a predefined threshold were identified as duplicates and subsequently removed, ensuring the dataset's integrity and reducing redundancy.
    
    \item {\tt Syntax Check:} The most crucial filtering step involved checking for syntax errors, as a dataset containing such errors would cause significant problems when training a model. To mitigate this, each file was processed using Icarus Verilog \cite{IcarusVerilog}, and files that failed due to syntax errors were discarded. Files containing other types of errors, such as missing imports or undefined references, were also labeled under "dependency issues" since these do not compromise the syntactic correctness of the code. Given that running Icarus Verilog is more computationally demanding than the earlier filtering steps, this step was performed last, ensuring it was applied to a reduced subset of files.
\end{itemize}

\subsubsection{Verilog code generation using commercial LLMs}
\begin{figure}
    \centering
    \includegraphics[width=1.0\columnwidth]{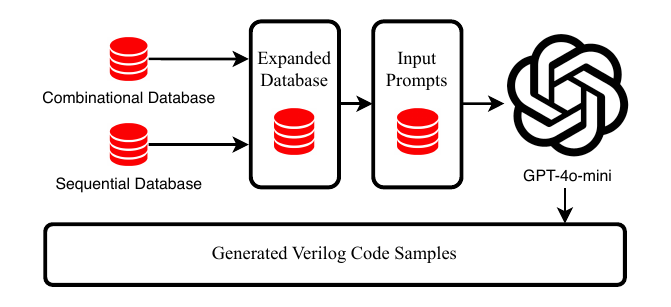}
    \caption{Process for generating Verilog code samples using commercial LLMs.}
    \label{fig:commercialGenerated}
\vspace*{-14pt}
\end{figure}

In addition to the collected code samples, we also generated a set of code samples using commercial LLMs, such as GPT-4o-mini. To guide this process, we first compiled a database of keywords and categorized them into combinational and sequential circuits, focusing on general hardware and Verilog design terms such as adders, multipliers, counters, FSMs, etc. For each keyword, we expanded the database by detailing specific variations, for instance, ripple carry adders or carry-save adders—this step was referred to as "expanded-keywords." Finally, for each expanded keyword, we crafted input prompts with detailed design descriptions and queried GPT-4o-mini 10 times for each prompt using different temperature values. Fig.~\ref{fig:commercialGenerated} depicts the overall process for the generating the Verilog code samples.

\subsubsection{Ranking and Complexity labels}
Each entry in the PyraNet dataset was evaluated and ranked using the GPT-4o-mini model. Rankings were assigned on a scale from 0 to 20, where 0 denotes the lowest score and 20 the highest. To facilitate the ranking process, we generated specific input prompts for each Verilog code sample in the dataset. The GPT-4o-mini model was instructed to score each code based on guidelines provided in a pre-prompt text, which detailed the scoring methodology. The scoring criteria focused on the overall Verilog coding style and the efficiency of the code. Fig.\ref{fig:RankingPromptSample} presents an example of a prompt along with its corresponding response.

\begin{figure}
    \centering
    \begin{tcolorbox}[colback=promptbg, colframe=promptbg]
    \small 
    \textbf{Prompt:} \\
    Act as a teacher and rank the quality of this Verilog code in scale of 0 to 20, with 0 being syntactically incorrect and 20 being a good Verilog code in terms of efficiency and coding style:
    \begin{lstlisting}[language=Verilog, frame=none, framesep=5pt, xleftmargin=5pt, xrightmargin=5pt]
    module halfAdder(
        input A,
        input B,
        output Sum,
        output Cout
        );
        
        assign Sum = A ^ B;
        assign Cout = A & B;
    endmodule
    \end{lstlisting}
    Just give me the score only.
    \end{tcolorbox}
    
    \begin{tcolorbox}[colback=responsebg, colframe=responsebg]
    \small 
    \textbf{Response:}
    \begin{lstlisting}[frame=none, framesep=5pt, xleftmargin=5pt, xrightmargin=5pt]
    Score: 20 out of 20.
    \end{lstlisting}
    \end{tcolorbox}
    
    \caption{Example of a prompt and the response used for ranking.}
    \label{fig:RankingPromptSample}
    \vspace*{-10pt}
\end{figure}

To assign a complexity level to each data sample in the PyraNet dataset, we closely followed the methodology presented in the MEV-LLM work \cite{Nadimi2024-MEV-LLM}. This approach allowed us to generate and categorize the dataset into four distinct tiers: {\tt Basic, Intermediate, Advanced}, and {\tt Expert}.

\subsubsection{Dataset Organization}
We organized the collected and generated Verilog code samples into six distinct layers:
\begin{itemize}
    \item [] {\tt \textbf{Layer 1}} comprises a total of $235$ data samples ranked 20 out of 20, representing the highest tier of Verilog codes. These samples are distinguished by their efficiency, coding style, and syntactic correctness. 

    \item [] {\tt \textbf{Layer 2}} includes $150,279$ data samples with rankings ranging from 19 to 15. 

    \item [] {\tt \textbf{Layer 3}} contains a total of $105,973$ data samples with rankings between 14 and 10. 

    \item [] {\tt \textbf{Layer 4}} includes $5,015$ data samples with rankings ranging from 9 to 5. 

    \item [] {\tt \textbf{Layer 5}} contains $275$ data samples with rankings between 4 and 1. 
    
    It is important to note that for Layers 1 through 5, we ensured representation across all complexity levels. Additionally, all samples in these layers successfully compiled without errors.

    \item [] {\tt \textbf{Layer 6}} contains a total of $430,461$ data samples with dependency issues that could not be compiled without errors or ranking of 0. 

\end{itemize}

Initially, we collected approximately 2.4 million code samples from open-source GitHub repositories and generated an additional 150,000 samples using GPT-4o-mini. Following the pre-processing phase, we obtained 692,238 code samples that were deduplicated, ranked, described, and compiled without syntax errors or having only dependency issues.

\subsection{PyraNet fine-tuning}
To effectively leverage the proposed PyraNet dataset, we have introduced a dedicated fine-tuning architecture. This architecture incorporates two key techniques: loss weighting and curriculum learning. In this section, we will first elaborate on the loss weighting approach, followed by a detailed explanation of the curriculum learning strategy.

\subsubsection{Loss Weighting}
As previously described, the PyraNet dataset is structured into six layers of data samples, each corresponding to different tiers. In our loss weighting technique, we fine-tune the model by assigning distinct loss weights to each tier within the dataset. Starting with the top layer—which contains highly ranked and high quality code samples—we set the loss weight to 1.0. This maximal weight ensures that entries in this layer have the most significant impact on the fine-tuning process. As we descend through the layers of the PyraNet dataset, the loss weights are progressively decreased. This reduction lessens the influence of data entries from lower tiers on the final fine-tuned model. Consequently, the model prioritizes learning from high-quality data samples while minimizing the effects of those with lower quality. For the layer of 2 to 6 the loss weights are 0.8, 0.6, 0.4, 0.2, and 0.1 respectively, as depicted in Fig.\ref{fig:overallFineTuningArchitecture}-b.

\subsubsection{Curriculum Learning}
As demonstrated in \cite{curriculumLearning}, employing a meaningful training sequence—beginning with simpler samples and progressing to more complex ones—can enhance model performance compared to standard training methods that use random sampling. In our proposed PyraNet dataset, we have assigned complexity levels to all code samples within each tier. Considering the benefits of curriculum learning, we propose initiating the fine-tuning process with the {\tt basic} complexity level for each tier, followed by {\tt intermediate}, {\tt advanced}, and finally {\tt expert} levels.

The fine-tuning process generally commences with the highest tier of the PyraNet dataset. Each tier within the PyraNet dataset is further categorized according to code complexity. Specifically, the process begins with data entries of {\tt basic} complexity within the top tier, followed by {\tt intermediate}, {\tt advanced}, and {\tt expert} complexity levels in sequence. This hierarchical structure is maintained across all tiers as the fine-tuning progresses downward through the dataset. 
By progressing through complexity levels in each tier, the model effectively leverages structured data to enhance its performance.
\vspace*{-4pt}

\section{Evaluation and Discussion}
\vspace*{-4pt}
\subsection{Baseline LLMs}
In this study, we aim to demonstrate the effectiveness of our proposed PyraNet dataset and the accompanying fine-tuning technique by conducting three distinct experiments and comparing the results with state-of-the-art models. 
We selected the CodeLlama-7b, CodeLlama-13b, and DeepSeek-Coder-7B models as our base models and fine-tuned them using the PyraNet dataset alongside our proposed fine-tuning methodology.
For evaluation, we employed the VerilogEval platform to assess the performance of the models across all experiments.

\vspace*{-6pt}
\subsection{Experiments}
To thoroughly evaluate the effectiveness of our proposed PyraNet dataset and the accompanying fine-tuning technique, we designed a series of three experiments. 
These experiments aim to systematically assess the individual and combined impacts of the dataset and the fine-tuning approach on model performance. 
The experimental setup is as follows:
\begin{itemize}
    \item {Baseline Experiment:} In the first experiment, we evaluated the CodeLlama and DeepSeek-Coder models by providing prompts without any fine-tuning. This established a baseline performance metric for comparison with subsequent experiments.
    \item {PyraNet-Only Fine-Tuning (PyraNet-Dataset):} In the second experiment, the CodeLlama and DeepSeek-Coder models were fine-tuned using the PyraNet dataset without applying the specialized fine-tuning approach. This allowed us to isolate and assess the improvements attributable solely to the PyraNet dataset.
    \item {Combined PyraNet dataset and PtraNet Fine-Tuning Approach (PyraNet-Architecture):} In the final experiment, we fine-tuned the CodeLlama and DeepSeek-Coder models using both the PyraNet dataset and our proposed fine-tuning technique. This combined approach was designed to evaluate the synergistic effects of the dataset and the fine-tuning methodology on model performance.
\end{itemize}
 
Through these experiments, we systematically evaluate the contributions of the PyraNet dataset and the fine-tuning technique, providing a comprehensive analysis of their impact compared to existing state-of-the-art models. The results are organized in Table~\ref{tab:results}.

\begin{table}[t]
    \centering
    \caption{PyraNet vs SOTA models on VerilogEval \cite{VerilogEval}}
    \newcolumntype{?}{!{\vrule width 2pt}}
    \label{tab:results}
    \resizebox{\columnwidth}{!}{%
    \begin{tabular}{c?c|c|c?c|c|c} 
        \multirow{2}{*}{Model}                              & \multicolumn{3}{|c?}{Verilog-Machine} & \multicolumn{3}{c}{Verilog-Human} \\ \cline{2-7} 
                                                            & $pass@1$        & $pass@5$        & $pass@10$       & $pass@1$        & $pass@5$        & $pass@10$        \\ \Xhline{5\arrayrulewidth}
        MG-Verilog-CodeLlama-7B \cite{mgverilog}            & 54.5            & 60              & 63              & -               & -               & -                \\ \hline
        RTLCoder-DeepSeek\cite{rtlcoder}                    & 61.2            & 76.5            & 81.8            & 41.6            & 50.1            & 53.4             \\ \hline
        OriGen-DeepSeek \cite{origen}                       & 74.1            & 82.4            & 85.7            & 54.4            & 60.1            & 64.2             \\ \Xhline{5\arrayrulewidth}
        codeLlama-7B-Instruct                               & 41.9            & 46.1            & 46.8            & 19.2            & 23.0            & 25.0             \\ \hline 
        codeLlama-7B-Instruct PyraNet-Dataset               & 58.0            & 62.9            & 67.8            & 44.2            & 50.0            & 55.7             \\ \hline
        codeLlama-7B-Instruct PyraNet-Achitecture           & 62.9            & 69.2            & 74.1            & 48.0            & 53.2            & 57.6             \\ \Xhline{5\arrayrulewidth}
        codeLlama-13B-Instruct                              & 48.6            & 54.5            & 57.3            & 32.6            & 35.8            & 39.1             \\ \hline 
        codeLlama-13B-Instruct PyraNet-Dataset              & 58.7            & 63.6            & 67.8            & 45.5            & 51.2            & 57.0             \\ \hline
        codeLlama-13B-Instruct PyraNet-Achitecture          & 69.2            & 73.4            & 78.3            & 50.6            & 55.1            & 58.3             \\ \Xhline{5\arrayrulewidth}
        DeepSeek-Coder-7B-Instruct-v1.5                     & 55.2            & 72.7            & 76.9            & 32.6            & 43.5            & 49.3             \\ \hline 
        DeepSeek-Coder-7B-Instruct-v1.5 PyraNet-Dataset     & 73.4            & 81.8            & 86.7            & 53.8            & 60.8            & 63.4             \\ \hline
        DeepSeek-Coder-7B-Instruct-v1.5 PyraNet-Achitecture & \textbf{77.6}   & \textbf{84.6}   & \textbf{89.5}   & \textbf{58.3}   & \textbf{62.8}   & \textbf{67.9}    \\ 
    \end{tabular}%
    }
\vspace*{-10pt}
\end{table}

\subsection{Fine-tuning Explained}
As outlined in the preceding sections, we fine-tuned two models: one using the PyraNet dataset without applying the proposed fine-tuning technique, and the other utilizing both the PyraNet dataset and our proposed fine-tuning method.

In the first experiment, we fine-tuned the CodeLlama-7B, CodeLlama-13B, and DeepSeek-Coder-7B models using each available (data, description) pair from the dataset. 
Specifically, the descriptions were used as inputs to the models, and the corresponding Verilog code served as the output. 
Throughout this experiment, the learning rate was maintained at $2e^{-4}$, and the loss weights were set to {$1.0$}.
The fine-tuning method utilizes the LoRa technique \cite{LoRa}, adhering to its standard training configurations.
The baseline LLM architectures chosen for this study, along with the fine-tuning parameters utilized, are succinctly summarized in Table~\ref{tab:fine-tuning_data}.

\begin{table}
    \centering
    \caption{Pre-trained LLM architectures and fine-tuning information}
    \label{tab:fine-tuning_data}
    \resizebox{\columnwidth}{!}{%
    \begin{tabular}{c|c|c|c|c|c|c} 
         Model                           & Layers & \# of Heads & Head Size            & Context Size             & learning rate              & \# of epochs  \\ \hline
         CodeLlama-7b-Instruct           & 32     & 32          & \multirow{3}{*}{128} & \multirow{2}{*}{100,000} & \multirow{3}{*}{$2e^{-4}$} &  \multirow{3}{*}{$1,2,3$} \\ 
         CodeLlama-13b-Instruct          & 40     & 40          &                      &                          &                            &                           \\ 
         DeepSeek-Coder-7B-Instruct-v1.5 & 30     & 30          &                      &  4000                    &                            &                           \\ 
    \end{tabular}%
    }
\vspace*{-14pt}
\end{table}

For the second experiment, we employed a hierarchical fine-tuning approach. We began by fine-tuning the aforementioned models with data entries from the top layer of the PyraNet dataset. This phase consisted of four sequential fine-tuning steps, starting with data entries labeled {\tt Basic}, followed by those labeled {\tt Intermediate}, {\tt Advanced}, and {\tt Expert}. In each step, descriptions were used as inputs and the corresponding Verilog codes as outputs, effectively implementing a curriculum learning strategy. During this initial fine-tuning phase, the loss weight was set to {$1.0$}, reflecting the use of the highest quality data available.

In subsequent rounds of fine-tuning, we progressed to lower layers of the PyraNet dataset, continuing the curriculum learning approach for each layer. The primary modification in these rounds was the adjustment of the loss weights, as illustrated in Fig.\ref{fig:overallFineTuningArchitecture}-b. By conducting multiple fine-tuning iterations with varying loss weights, we implemented the loss weighting approach. The fine-tuning process concluded upon completing the lowest layer of the PyraNet dataset. It is important to note that the learning rate was maintained at a constant value of $2e^{-4}$ for each round of fine-tuning throughout this experiment as well.
This structured methodology allowed us to systematically assess the impact of both the hierarchical fine-tuning and loss weighting techniques on model performance, ensuring that the models effectively leveraged high-quality data while appropriately incorporating varying levels of complexity from the dataset.
\vspace*{-6pt}

\subsection{Comparison and Explanation of the Results}
\vspace*{-2pt}

To ensure a fair evaluation, the proposed PyraNet fine-tuning approach was applied to both the CodeLlama model, for comparison with MG-Verilog \cite{mgverilog}, and the DeepSeek-Coder model, for comparisons with RTL-Coder \cite{rtlcoder} and OriGen \cite{origen}. 
The results in Table \ref{tab:results} indicate that the proposed model achieved up to a $32.6\%$ improvement in the $pass@k$ metric over the CodeLlama baseline models and up to $25.7\%$ over the DeepSeek-Coder baseline model.
When compared to state-of-the-art models, the proposed approach demonstrated an $11.1\%$ improvement in the $pass@k$ metric on the Verilog-Machine dataset compared to MG-Verilog. 
For the Verilog-Human, the proposed model achieved improvements of up to $16.7\%$ against RTL-Coder and $3.9\%$ against OriGen, both of which are based on fine-tuning with DeepSeek-Coder.

It is worth noting that OriGen incorporates a Self-Reflection loop, which includes an additional round of error correction—a feature not considered in this work due to time constraints.
This accounts for the relatively smaller improvements over OriGen. 
Nevertheless, integrating OriGen's Self-Reflection loop with inference using the models fine-tuned with the proposed PyraNet architecture is expected to yield even greater performance gains, which will be performed in the future. 
All the gains achieved by the proposed PyraNet dataset and architecture are illustrated in Table~\ref{tab:gains}.

\begin{table}[t]
    \centering
    \caption{PyraNet Gains vs Baseline Model and SOTA}
    \newcolumntype{?}{!{\vrule width 2pt}}
    \label{tab:gains}
    \resizebox{\columnwidth}{!}{%
    \begin{tabular}{c|c?c|c|c?c|c|c} 
        \multirow{2}{*}{Model}                                                                           & \multirow{2}{*}{VS} & \multicolumn{3}{|c?}{Verilog-Machine} & \multicolumn{3}{c}{Verilog-Human} \\ \cline{3-8} 
                                                                                                         &                     & $pass@1$ & $pass@5$ & $pass@10$ & $pass@1$ & $pass@5$ & $pass@10$ \\ \Xhline{5\arrayrulewidth}
        codeLlama-7B-Instruct                                                                            & vs Baseline         & 16.1     & 16.8      & 21	     & 25       & 27       & 30.7      \\ \cline{2-8}
        PyraNet-Dataset                                                                                  & vs MG-Verilog       & 3.5	  & 2.9	      & 4.8      & -        & -        & -         \\ \Xhline{5\arrayrulewidth}
        codeLlama-7B-Instruct                                                                            & vs Baseline         & 21	      & 23.1	  & 27.3	 & 28.8	    & 30.2	   & 32.6      \\ \cline{2-8}
        PyraNet-Achitecture                                                                              & vs MG-Verilog       & 8.4      & 9.2	      & 11.1     & -        & -        & -         \\ \Xhline{5\arrayrulewidth}
        codeLlama-13B-Instruct                                                                           & vs Baseline         & 10.1     &	9.1       &	10.5	 & 12.9	    & 15.4     & 17.9      \\ \cline{2-8}
        PyraNet-Dataset                                                                                  & vs MG-Verilog       & 4.2      & 3.6	      & 4.8      & -        & -        & -         \\ \Xhline{5\arrayrulewidth}
        codeLlama-13B-Instruct                                                                           & vs Baseline         & 20.6	  & 18.9	  & 21	     & 18	    & 19.3	   & 19.2      \\ \cline{2-8}
        PyraNet-Achitecture                                                                              & vs MG-Verilog       & 14.7	  & 13.4	  & 15.3     & -        & -        & -         \\ \Xhline{5\arrayrulewidth}
        \multirow{3}{*}{\parbox{4cm}{\centering DeepSeek-Coder-7B-Instruct-v1.5 \\ PyraNet-Dataset}}     & vs Baseline         & 18.2	  & 9.1	      & 9.8	     & 21.2	    & 17.3	   & 14.1      \\ \cline{2-8}
                                                                                                         & vs RTL-Coder        & 12.2	 & 5.3	     & 4.9	    & 12.2	   & 10.7	  & 10        \\ \cline{2-8}
                                                                                                         & vs OriGen           & -0.7	 & -0.6	     & 1	    & -0.6	   & 0.7	  & -0.8      \\ \Xhline{5\arrayrulewidth}
        \multirow{3}{*}{\parbox{4cm}{\centering DeepSeek-Coder-7B-Instruct-v1.5 \\ PyraNet-Achitecture}} & vs Baseline         & 22.4	  & 11.9	  & 12.6	 & 25.7	    & 19.3	   & 18.6      \\ \cline{2-8}
                                                                                                         & vs RTL-Coder        & 16.4	 & 8.1	     & 7.7	    & 16.7	   & 12.7	  & 14.5      \\ \cline{2-8}
                                                                                                         & vs OriGen           & 3.5	     & 2.2	     & 3.8	    & 3.9	   & 2.7	  & 3.7       \\
    \end{tabular}%
    }
\vspace*{-10pt}
\end{table}

\vspace*{-4pt}

\subsection{Dataset Quality Verification}
\vspace*{-2pt}
As emphasized in \cite{VerilogEval}, the integrity of the compiled dataset is crucial for the effectiveness of the fine-tuning process. 
Due to challenges in verifying each description and additional information incorporated into the PyraNet dataset via the GPT-4o-mini model, we adopted the verification methodology employed by the researchers in \cite{VerilogEval} to validate the labels generated by the GPT-4o-mini model.
To assess the quality of the dataset, we randomly shuffled the codes, descriptions, and ranking information among the data entries, thereby creating mismatched sets of codes, descriptions, and rankings within each row of the PyraNet dataset. 
We then proceeded to fine-tune the model using this intentionally distorted dataset.
Table~\ref{tab:datasetQuality} presents the results of fine-tuning the CodeLlama-7B model with both the flawed and accurate datasets. 
The findings clearly demonstrate that the model's code generation capabilities are significantly compromised when trained on the erroneous dataset. 
This outcome verifies the accuracy of the labels and underscores the critical importance of dataset quality in the fine-tuning process.
Furthermore, the results demonstrate that the accuracy of the fine-tuned model decreases when trained on the erroneous dataset compared to training solely on the PyraNet dataset without the proposed fine-tuning technique. Consequently, we decided to forgo the fine-tuning process using the proposed method, as it is evident that an erroneous dataset would similarly impair the effectiveness of the fine-tuning technique and result in reduced model performance.
\vspace*{-6pt}
\begin{table}
    \centering
    \caption{Results for erroneous dataset}
    \newcolumntype{?}{!{\vrule width 2pt}}
    \label{tab:datasetQuality}
    \resizebox{\columnwidth}{!}{%
    \begin{tabular}{c?c|c|c?c|c|c} 
        \multirow{2}{*}{Model}                   & \multicolumn{3}{|c?}{Verilog-Machine} & \multicolumn{3}{c}{Verilog-Human} \\ \cline{2-7} 
                                                 & $pass@1$ & $pass@5$ & $pass@10$ & $pass@1$ & $pass@5$  & $pass@10$  \\ \Xhline{5\arrayrulewidth}
        CodeLlama-7B with erroneous dataset      & 43.3     & 48.2     & 50.3      & 21.1     & 25.6      & 28.8       \\ \hline
        CodeLlama-7B with correct dataset        & 58.0     & 62.9     & 67.8      & 44.2     & 50.0      & 55.7       \\ 
    \end{tabular}%
    }
\vspace*{-10pt}
\end{table}

\section{Conclusion and Future Works}

In this study, we present a novel structured dataset, PyraNet, alongside its associated fine-tuning architecture. This architecture addresses the limitations of previous studies by achieving improved performance across diverse Verilog code types. The PyraNet fine-tuning framework employs advanced techniques such as loss weighting and curriculum learning to mitigate the challenges posed by the available datasets. Evaluation results reveal that the PyraNet dataset and its proposed architecture achieve the anticipated benefits, demonstrating up to a $32.6\%$ improvement in the $pass@k$ metric. We suggest that there is substantial scope for further advancements, both through the development of more comprehensive and detailed datasets and by exploring alternative fine-tuning architectures.
\section*{Acknowledgment}

This work is supported by the National Science Foundation under Grant No. 2434247. 
Any opinions, findings, and conclusions or recommendations expressed in this material are those of the author(s) and do not necessarily reflect the views of the funding agencies.

\vspace*{-10pt}

\clearpage

\end{document}